\documentstyle[12pt,epsf]{article}

\begin{document}

\newcommand{\myprime}{^\prime}
\newcommand{\grad}{\nabla}
\newcommand{\mass}{{\cal M}}
\newcommand{\lapse}{N^{t}}
\newcommand{\shift}{N^{r}}

\title{\begin{flushright}
\begin{small}
hep-th/9604189,  PUPT-1617  
\end{small} 
\end{flushright}
On D-Branes and Black Holes  in Four Dimensions}
\author{Vijay Balasubramanian and  Finn
Larsen\thanks{{vijayb,larsen}@puhep1.princeton.edu}\\    
       {\it Joseph Henry Laboratories, Princeton University,}\\
         {\em Princeton,   NJ 08544} }
\date{\today}                                 
\maketitle                                 

\begin{abstract}
We find extremal four dimensional black holes with finite
area constructed entirely from intersecting D-branes.  We argue that
the microscopic degeneracy of these configurations agrees with the
Bekenstein-Hawking entropy formula.  The absence of solitonic objects
in these configurations may make them useful for dynamical studies of
black holes.
\end{abstract}

\section{Introduction}
In recent months there has been considerable progress in accounting
for the microscopic degeneracy of black holes using the explicit
conformal field theories provided by D-branes~\cite{polch95a}.
Examples of black holes in four and five dimensions have been
constructed for which the degeneracy of microscopic D-brane states
matches the Bekenstein-Hawking
entropy~\cite{strom96a} -~\cite{bmpv}. 
The four dimensional configurations considered
so far contain some combination of D-branes and solitons and  seem
unwieldy for investigation of dynamical phenomena.
D--branes are presently understood far better that solitons in 
general; so it may be an advantage to construct four dimensional
black holes entirely from D--branes. In this paper we present
some aesthetically pleasing configurations that realize this,
and discuss their microscopic entropy.

The paper is organised as follows.  In Section~\ref{sec:branehole} we
discuss classical black hole solutions to the IIA and IIB
supergravities and find their area.  Regularity conditions on the
classical p-brane solutions lead us to the most general construction
of black holes from four intersecting D--branes of Type II on $T^6$.
In Section~\ref{sec:entropy} we discuss the microscopic degeneracy of
these systems of branes and their relationship to intersecting brane
configurations of M-theory.

\section{Brane Surgery} 
\label{sec:branehole}

We are interested in configurations of D--branes wrapped on $T^6$
whose four dimensional manifestations are regular extremal black
holes.  To achieve this, the dilaton and the moduli (the metric of the
compactified dimensions) must be finite at the horizon in order to
avoid large corrections to the low energy solution. It is also
important that the black hole preserves some supersymmetry so that
quantum corrections are well controlled. We will consider
systems of four intersecting D--branes.  This is the smallest
number for which all these requirements can be met.    The configurations 
in this paper are dual to the black holes of Cveti\v{c}
and Youm~\cite{cveticyoum}. 

The string metric and the dilaton of a single D--brane are~\cite{strom91}:
\begin{eqnarray}
ds^2 &=& F^{-1/2} \left ( - dt^2
+dx_1^2 + \ldots + dx_p^2 \right )+
F^{1/2} \left ( dx_{p+1}^2 + \ldots + dx_9^2 \right ) \nonumber \\
e^{-2 \phi } &=& F^{(p-3)/2}
\label{metric}
\end{eqnarray}
The profile function $F$ is a solution of the Laplace equation, so
that after compactification to 4 dimensions $F=1+ \frac{q}{r}$ for all
branes.  Here $r^2=x_7^2+x_8^2+x_9^2$ is the radius squared in the
uncompactified dimensions.  The metric and the dilaton for the
intersecting branes of interest here can be found by multiplying the
profile functions for each of the branes~\cite{tseytlin96,gauntlett96}. The
complete solution obtained this way has a horizon at $r=0$.

The black hole must have a finite dilaton at the horizon.  For a
single D--brane the dilaton is finite at the horizon only for $p=3$.
For multiple intersecting D--branes the p's must be chosen in such a
way that the product of the profile functions tends to a finite value
at the horizon.  For four intersecting D-branes it is necessary and
sufficient that the p's add up to 12.  For example, one 6D-brane and
three 2D-branes satisfy this criterion.\footnote{We do not consider
systems of 2 and 3 intersecting branes because they cannot satisfy all
three regularity conditions described in this section.}

Regular black holes must also have finite moduli at the horizon.  The
profile function $F$ diverges in the same way for all branes, and the
dimensions parallel and transverse to a brane are multiplied by
$F^{-1/2}$ and $F^{1/2}$ respectively.  So each compact dimension must
be perpendicular to as many branes as it is parallel to.  Four D--branes
can satisfy this criterion if each dimension is in the worldvolume of
exactly two D--branes.  For example, four 3-branes wrapped around the
cycles (123), (345), (146) and (256) satisfy both the finite dilaton
and finite moduli conditions.

Each D--brane imposes a boundary condition that relates the
components of spacetime spinors. This identifies the two
supersymmetries in 10 dimensions in a way that depends on the
orientation of the D--brane.  When two D--branes are present we get
two boundary counditions.   The first identifies the two supersymmetries
and the second leads to the projection condition:
\begin{equation}
 {\cal Q} = \pm \Gamma {\cal Q} ~~;~~~~
\Gamma = \Gamma_{a_1}\cdots\Gamma_{a_n}
\label{projection}
\end{equation}
on the remaining generator in 10 dimensions. The sign in the relation
parametrises the distinction between branes and antibranes and is
purely conventional.  The indices $a_i$ run over values that are
present for one of the branes but not for the other ( i.e. (ND)
directions ).  The condition that this is indeed a projection
($\Gamma^2 =1$) so that there is some surviving supersymmetry,
requires that a p-brane intersecting a q-brane along $k$ dimensions
satisfies $p + q - 2k = 0 \bmod 4$.

In addition to this condition on each pair, we must consider the
complete system of 4-branes.  One of the branes identifies the two
supersymmetries and the other three impose three projection conditions
of the type in Eq.~\ref{projection}.
Without loss of generality, the projection operators for four
intersecting branes can be taken to
be $ \pm\Gamma_1 \Gamma_2 \Gamma_3 \Gamma_4$, $\pm \Gamma_1 \Gamma_2
\Gamma_5 \Gamma_6$, and $\mp \Gamma_3 \Gamma_4 \Gamma_5
\Gamma_6$. These matrices commute so that all the different conditions are
compatible.  Each of the 4 branes imposes one condition and therefore
breaks $1/2$ of the supersymmetry. However, one of these conditions is
redundant - two of the three projection operators multiply to give the
third, so the complete state only breaks $1/8$ of the supersymmetry.

Note that one of the three projections has a sign that is determined
by the others. Therefore the signs in Eq.~\ref{projection} can not be
chosen independently for each of the 4 D--branes. There are $2^4=16$
distinct assignments of orientation for the D--branes but only 8 of
them lead to 1/8 supersymmetric configurations.

As a final condition recall that a given configuration can only 
contain even branes in IIA or
odd branes in IIB.  All the conditions taken together are very
restrictive.  For example, assume that the configuration includes a
6--brane wrapped around the $T^6$.  Then the supersymmetry condition
requires that only 2--branes or 6-branes may be intersected with the
6--brane.  The dilaton condition dictates that there must be three
2-branes, and the moduli condition requires that the 2--branes should
lie in orthogonal dimensions.  In this way we get a complete
classification of possible configurations of intersecting D-branes
that form regular black holes with finite area in 4 dimensions.


Let $1,\cdots ,6$ denote the dimensions of the $T^6$ and use p--tuples
for worldvolume coordinates of a D--brane. For example, (12) is a
2D--brane aligned along the first and second compactified dimension.
In this notation, the moduli are stabilised when each of the six digits
enters exactly twice.  The regularity conditions discussed above yield
the configurations:
\begin{eqnarray}
(123),(345),(146),(256) \nonumber \\
(1234),(3456),(1256),() \nonumber \\
(1234),(3456),(12),(56) \nonumber \\
(12345), (126), (346), (5) \nonumber \\
(123456),(12),(34),(56) 
\label{branes}
\end{eqnarray}
The second configuration was mentioned in \cite{cveticsen} .

Recall that T--duality along a given direction acts on the branes by
adding the corresponding index if it is not there already, and
removing it, if it is.  Then it is easy to see that the examples given
are in fact T--dual to each other. In this sense, there is a unique
regular supersymmetric black hole in 4 dimensions that can be made out
of four D--branes. It is quite remarkable that such symmetric
configurations exist at all.  We find it intriguing that this
possibility is special to four dimensions.

The configuration consisting of 4 3D--branes is particularly
symmetric. The metric is:
\begin{eqnarray}
ds^2 &=& (F_1 F_2 F_3 F_4)^{-1/2} ( -dt^2)
+(F_1 F_2 F_3 F_4)^{1/2} (dx_7^2 + dx_8^2 + dx_9^2 )  \nonumber \\
 &+& [ \left( \frac{F_1 F_2}{F_3 F_4} \right)^{1/2} dx_1^2
+ \left( \frac{F_1 F_3}{F_2 F_4} \right)^{1/2} dx_2^2
+ \left( \frac{F_1 F_4}{F_2 F_3} \right)^{1/2} dx_3^2
+ \left( \frac{F_2 F_3}{F_1 F_4} \right)^{1/2} dx_4^2 \nonumber \\
& +& \left( \frac{F_2 F_4}{F_1 F_3} \right)^{1/2} dx_5^2
+ \left( \frac{F_3 F_4}{F_1 F_2} \right)^{1/2} dx_6^2
]
\label{branemetric}
\end{eqnarray}
For 3D--branes the dilaton is constant so there is no need to 
distinguish between Einstein metric and string metric. 
The four dimensional area is
\begin{equation}
A = 4\pi \sqrt{q_1 ~q_2 ~q_3 ~q_4}
\label{area}
\end{equation}
where the $q$'s are the charges of the four different branes.

For this configuration, the volume of the compact space is independent
of radial distance from the horizon.
 Therefore the area of the black hole in the ten dimensional theory is
given by Eq.~\ref{area}, multiplied by the constant internal volume.
Also recall that $G_N$ in the four dimensional theory differs from the
one appropriate to the ten dimensional theory by the volume of the
compactified space, measured at infinity. Therefore the entropy of
this black hole:
\begin{equation}
S = \frac{A}{4G_N} 
\end{equation}
is the same when $A$ and $G_N$ are understood either from the ten
dimensional perspective or in the compactified theory.

The form of the entropy that is suitable for comparison with microscopic 
considerations is the expression in terms of integer quanta of 
the D--brane charges. The appropriate formula is
\begin{equation}
S= 2\pi\sqrt{Q1 ~Q2 ~Q3 ~Q4}
\label{bhentropy}
\end{equation}
Here the notation $Qi$ denotes the integer quanta of the four branes.
This can be shown from saturation of Dirac's quantization
condition. Despite this topological origin the derivation known at
present involves explicit consideration of the normalization of
charges~\cite{kallosh96a}.

The entropy is a pure number independent of both the moduli of the
torus and the string coupling.  This is a necessary condition for any
counting to work \cite{larsen95,kallosh96b}.  Moreover, the Einstein
area is invariant under T--duality so that same entropy formula is
relevant to all the configurations considered here. In fact, the
entropy formula can be uniquely extended to the full U--duality group
$E(7)$ \cite{kallosh96a} .

\section{Microscopic Entropy}
\label{sec:entropy}
The regularity requirements imposed in the previous section allowed us
to identify collections of D-branes that act as black holes with
finite area.  The specific classical solution written down in
Eq.~\ref{branemetric} was chosen to have translational symmetry in the
compact dimensions.  Such choices regarding the details of the
configuration in the internal space are not unique - there is a
spectrum of degenerate microscopic states consistent with the choice
of charges measurable at infinity.\footnote{For a discussion of some
of the spacetime aspects of this point see~\cite{larsen96}.}  In this
section we examine the microscopic D--brane derivation of the
corresponding entropy.

\subsection{Intersecting D--branes}
\label{sec:source}
Let us briefly recall the main features of some successful D--brane
state counts.  As an example, consider $N$ 0-branes on $M$ parallel,
intersecting 4-branes.  The 0-branes are described as instantons of the
4-brane SU(M) world-volume
theory~\cite{strom96a,vafa95a,vafa95b,douglas95a}.  Consequently, 
each 0-brane comes equipped with $4M$ degrees of freedom associated with
orientations of the instantons of SU(M).  A given classical
configuration breaks $4NM$ symmetries because it defines a point in
the moduli space.  The corresponding Goldstone modes (and their
superpartners) are the relevant elementary excitations leading to the
entropy. They are in one-to-one correspondence with the strings on the
intersection manifold that run between branes and bind them
together~\cite{douglas95a}.  This correspondence is one of the
attractive features of D--branes that allows a simple counting of the
Goldstone modes.

We are interested in multiple intersecting D-branes that are not
parallel.  We expect a non-trivial interacting theory on the
intersection manifold and therefore it appears to be difficult to
count the number of zero modes explicitly.  Presumably the
intersecting configurations are described by certain classical
excitations of the branes can be ``oriented'' relative to each other,
and these ``orientations'' are described by the zero modes of the
string condensate binding the branes together.

\subsection{4-4-4-0}
\label{sec4440}
The Type IIA 4-4-4-0 configuration has Q1, Q2 and Q3 4-branes wrapped
around the (1234), (3456) and (1256) cycles of $T^6$.  Let us first
consider the special case $Q1=Q2=Q3 =1$.  The three 4-branes on $T^6$
intersect at exactly one point.  Now bind 0-branes to this mutual
intersection point by turning on a background of strings running
between the 0-branes and each of the 4-branes.  This can be thought of
as a condensate of strings that describes orientation degrees of
freedom of the 0-brane on the intersecting 4-branes.  A 0--brane bound
to the intersection point of three 4-branes can carry any integer
multiple of the 0-brane RR quantum.  These multiply charged 0-brane
states arise as fundamental zero-branes bound together by giving an
expectation value to the massless strings running between them.  (This
picture follows naturally from M-theory as we will describe in the
next section.)  Such 0-brane bound states should be counted as
separate sectors of the Hilbert space in counting the degeneracy of a
configuration.\footnote{The twisted sectors of the moduli space of
identical 0-branes on 4-branes (\cite{vafa95a,vafa95b}) can be
interpreted as precisely such bound states, and they are indeed
counted as separate states in the Hilbert space.}

Let us suppose that a 0-brane attached to the intersecting 4-branes
has $k$ bosonic orientation degrees of freedom. Then the unbroken
supersymmetry implies an equal number of fermionic orientations.
Define $a_n^{\dagger\mu}$ ( $\mu = {1\cdots k}$) to be an operator
that creates a charge $n$ 0-brane with bosonic orientation $\mu$ at
the intersection point.  Let $b_n^{\dagger\mu}$ be the corresponding
fermionic operator.  Then the zero-brane charge operator is $Q =
\sum_{n=1}^{\infty} n a_n^{\dagger\mu} a_n^{\mu} +
\sum_{n=1}^{\infty} n b_n^{\dagger\mu} b_n^{\mu}$.  The generating
function of the degeneracy of states with charge $n$ is then given by
the familiar formula:
\begin{equation}
\sum_n d(n) q^n = \frac{\prod_n (1 + q^n)^{k}}{\prod_n (1 - q^n)^k}
\end{equation}
The degeneracy of states with three 4-branes and total 0-brane charge
Q4 is $d(Q4) = \exp{2\pi \sqrt{(k/4) \,Q4}}$ for $Q4\gg 1$

Let us return to the general case where Q1, Q2 and Q3 are greater than
unity.  Then each of the three varieties of 4--branes consists of a
number of branes that are parallel but need not coincide. Separating the
branes gives $Q1\,Q2\,Q3$ distinct spacetime points where there are
three intersecting branes.  Attach a total 0-brane charge of $Q4$ to
these intersection points.  Each of the $Q1 \, Q2 \, Q3$ intersection
points contributes $k$ bosonic and $k$ fermionic modes.  As before
introduce creation operators describing the charge $n$ 0-brane states,
except that the index $\mu$ on the operators $a$ and $b$ now runs
between $1$ and $k\,Q1\,Q2\,Q3$.  The degeneracy of states becomes
$d(Q4) = \exp{2\pi \sqrt{(k/4) \,Q1\, Q2\, Q3\, Q4}}$ for large $Q4$
and the entropy is
\begin{equation}
S =  \ln d = 2\pi \sqrt{(k/4) \, Q1 \, Q2 \, Q3 \, Q4}
\label{eq:braneentropy}
\end{equation}
This agrees with the Bekenstein-Hawking entropy calculated in
Section~\ref{sec:branehole} when $k = 4$.  We therefore expect that a
0-brane bound to the intersection point of three 4-branes has 4
bosonic and 4 fermion zero modes.

It can be understood heuristically why $k=4$: two 4--branes break
$3/4$ of the 32 real spacetime supersymmetries; so the worldvolume
theory on each intersection manifold respects the remaining 8. Half of
these are broken when another D--brane is attached and the Goldstone
modes associated with this final breaking can condense to bind the
4--branes. There is a separate condensate at each intersection point
that comprises 4 fermionic modes and, by virtue of the unbroken
worldvolume supersymmetry, 4 bosonic modes. We will motivate in the
following section the expectation that excitations of the condensate
can carry 0--brane RR charge.  As shown in Section 2, the 0--brane RR
charge does not break any additional supersymmetry; so it is expected
that $k=4$.
\footnote{Collective coordinates were attributed to  
supersymmetries broken by one of the three 4--branes but respected by
the others.  A more democratic treatment might have suggested a
three-fold duplication of these Goldstone modes, but we presume that
the correct construction effectively identifies these copies.}

This argument is essentially that of \cite{dvv96a} . It also works 
for the five dimensional black holes described by a 1-brane and a 5-brane:
the worldvolume theory of the 5-brane has 16 supersymmetries,
the 1-brane breaks $1/2$ of them, and 8 Goldstone modes appear.
Excited states of the binding condensate can carry momentum but,
for this to preserve some of the remaining supersymmetry, it
is only half of the modes that effectively participate. This
leaves $k=4$ orientations.

\subsection{M-theory and 4d Black Holes}
\label{sec:M}
A number of issues regarding the state-counting for the 4-4-4-0
configuration are clarified from an M-theory perspective.  Compactify
M-theory on a circle and let 0 denote the compact dimension.  Then the
three 4-branes arise as compactifications of the 5-branes (01234),
(01256) and (03456).  The RR 1-form gauge field is the Kaluza-Klein
gauge field associated with the 11-dimensional metric.  The 0-brane
RR charge descends from the quantized momentum along the compact
dimension and leads  to  multiply charged 0-branes
in 10 dimensions.     These multiply charged states are not just
two juxtaposed fundamental 0-branes, because a mode carrying momentum
$2n$ in the 0 direction is physically distinct from two modes carrying
momentum $n$ each.  These states were used in the previous section.

In the 4-4-4-0 counting the enormous degeneracy arose because of the
many mutual intersection points.  This simplification is only suitable
for the limit where Q4 is much bigger than Q1, Q2 and Q3.  Indeed, the
4--branes may bind, just like the 0--branes; so we may expect states
with multiple 4--brane charge that would reduce the number of
intersection points and threaten the state counting.  The problem is
resolved by noting that there are two ways to obtain a bound state of
4-branes from M-theory.  The naive way is to bind $Q1$ 5-branes together
and dimensionally reduce to get a bound state of $Q1$ 4-branes.  
Alternatively, we can wind a 5-brane $Q1$ times around the 0 direction
and dimensionally reduce to get a bound state of 4-branes.  In this
case, the effective length of the circle on which the Kaluza-Klein
modes live is multipled by $Q1$ leading to 0-brane charges quantized
in units of $1/Q1$.  Admitting these fractionally charged 0-branes
gives the correct state count~\cite{mathur96,susskind96}.  In our
case, this mechanism has to employed for each of the three
intersecting varieties of 5-branes, leading to Kaluza-Klein modes
quantized in units of $1/(Q1\,Q2\,Q3)$ (for relatively prime $Q1$,
$Q2$ and $Q3$).\footnote{Similar arguments have also appeared recently
in~\cite{igor96a}.}  When all four charges are big~\cite{susskind96}
it is essential to consider these additional states in order to match
the Bekenstein-Hawking formula.  The ``fractional charge'' required
here is the D-brane analogue of the ``tension renormalization''
of~\cite{larsen95}.

M-theory also clarifies the nature of the string condensate binding
the 4-4-4-0 configuration together. M-theory 5-branes interact via
exchange of membranes with boundaries on the
5-branes~\cite{Mbeckers,Mstrom,Mtownsend}, and quantization of the
boundary states yields the 5-brane low-energy effective
theory~\cite{dvv96a,dvv96b}.  This suggests that collapsed membranes
live on the intersection manifold of our intersecting 5-branes and
bind them together~\cite{igor96a}.  Dimensional reduction of these
collapsed membranes would give the desired string condensate and the
momentum of the membranes turns into 0-brane charge.  It is suggested
in~\cite{igor96a} that the collapsed membranes act like a single
self-dual string in 6 dimensions with 4 bosonic and fermionic
zero modes.  This would give $k=4$ in the 4-4-4-0 counting, but the
rules for state counting in M-theory are not yet firmly
established~\cite{schwarz96a} and more work is necessary on this
point.

\section{Comments on Other Configurations}

\subsection{3-3-3-3}
\label{sec:3333}

The 4-4-4-0 configuration is T-dual to the more symmetric IIB 3-3-3-3
configuration.  There is a heuristic counting of
states for this configuration  that is exactly parallel to
the IIA setup.  Once again, take Q1, Q2 and Q3 3-branes wrapped
around the (123), (345) and (146) cycles of $T^6$ respectively.  Then
there are $Q1\, Q2\, Q3$ distinct spacetime points where three branes
intersect.  We attach 3-branes wrapped around the (256) cycle to these
points.  By T-duality from the 4-4-4-0 configuration we expect that
these (256) branes come in multiply charged varieties.  We suppose as
before that there are $k$ bosonic and $k$ fermionic ``orientations''
of the fourth species of 3-brane that arise as collective excitations
of the strings running between the branes. Then the combinatorics of
the previous section goes through unchanged giving an entropy of $S =
2\pi \sqrt{(k/4) \, Q1 \, Q2 \, Q3 \, Q4} $.

The existence of the 3-3-3-3 configuration makes it manifest
that the entropy must be symmetric in all charges. It is disappointing
that it appears very difficult to extract any additional insight
from this remarkable configuration.



\subsection{2-2-2-6}
\label{sec:2226}
The 2-2-2-6 black hole in Type IIA is particularly suggestive.  In
this case the Q4 6-branes are completely wrapped around $T^6$ so it is
not possible to separate them as we did in the 3-3-3-3 and 4-4-4-0
cases.  Now the $\int C^{(3)} \wedge F \wedge F$ coupling between the RR
3-form and the world-brane gauge field strength implies that each
2-brane separately looks like a an instanton of the SU(Q4) 6-brane
gauge theory~\cite{douglas95a,polch96a}.  Therefore the intersecting
2-branes on a 6-brane should correspond to a classical solution of 6
dimensional Euclidean gauge theory with the property that projection
onto the (1234), (3456) or (1256) cycles yields a standard
four-dimensional instanton.  It would very interesting to count the
degeneracy of this intersecting brane system by counting the
orientation degrees of freedom of these six-dimensional
``instantons''.

\section{Conclusion}
\label{sec:conc}
We have presented four dimensional extremal black hole configurations
with finite area that are composed entirely out of D-branes.  We
argued that the entropy of these configurations naturally scaled
with the charges to match the Bekenstein-Hawking entropy formula.  
We gave heuristic arguments that fixed the overall
coefficient in the entropy, but much more needs to be done to
establish the rules for this counting.  The Type IIB configuration of
four intersecting 3-branes is particularly interesting because the
complete symmetry between the various branes explicitly reflects the
symmetry of the entropy formula.  A possible advantage of the black
hole configurations presented here is that they do not contain any
solitonic objects.  Consequently it should be easier to conduct
scattering experiments from them than from the previously constructed
black hole configurations.

\section{Acknowledgements}
\label{sec:ack}
We would like to thank Curtis Callan, Rajesh Gopakumar,
Igor Klebanov, Juan Maldacena, Ashoke Sen, and Eric Sharpe for useful
conversations. We had several fruitful discussions with Mirjam
Cveti\v{c} who suggested that black holes can be made of D--branes
alone. FL also gratefully acknowledges an ongoing
collaboration with Frank Wilczek on closely related issues.  While
this paper was in preparation we received~\cite{igor96a} which has
some overlap with the material presented here.  VB is supported in
part by DOE grant DE-FG02-91ER40671.


\end{document}